\documentclass[journal,12pt,onecolumn,draftclsnofoot]{IEEEtran}

\usepackage{color}
\usepackage{multirow}
\usepackage{graphicx}
\usepackage{epstopdf}
\usepackage[cmex10]{amsmath}
\usepackage{bm} 
\usepackage{amssymb}

\usepackage{multirow}
\usepackage{amsthm}
\usepackage{cite}
\usepackage{balance}
\usepackage{acronym} 
\usepackage{algorithm, algcompatible}

\newcommand{\diag}{\mathop{\rm diag}}

\algnewcommand\INPUT{\item[\textbf{Input:}]}
\algnewcommand\OUTPUT{\item[\textbf{Output:}]}

\allowdisplaybreaks

\begin{document}

\title{Spatially Correlated multi-RIS Communication:\\The Effect of Inter-Operator Interference} 
\author{Nikolaos~I.~Miridakis,~\IEEEmembership{Senior Member,~IEEE}, and Panagiotis~A.~Karkazis 
\thanks{The authors are with the Department of Informatics and Computer Engineering, University of West Attica, Aegaleo 12243, Greece (e-mails: \{nikozm, p.karkazis\}@uniwa.gr).}
\thanks{The work of the authors is implemented in the framework of H.F.R.I call ``3rd Call for H.F.R.I.'s Research Projects to Support Faculty Members \& Researchers'' (H.F.R.I. Project Number: 23291). It is also partially supported by the SAFE-6G project funded by the Smart Networks and Services Joint Undertaking (SNS JU) under the European Union's Horizon Europe research and innovation programme under Grant Agreement No101139031.}
}


\maketitle

\begin{abstract}
A multi-operator wireless communication system is studied where each operator is equipped with a reconfigurable intelligent surface (RIS) to enhance its communication quality. RISs controlled by different operators affect the system performance of one another due to the inherently rapid phase shift adjustments that occur on an independent basis. The system performance of such a communication scenario is analytically studied for the practical case where spatial correlation occurs at RIS of arbitrary size. The proposed framework is quite general since it is analyzed under Nakagami-$m$ channel fading conditions. Finally, the derived analytical results are verified via numerical and simulation trials as well as some new and useful engineering outcomes are revealed.  
\end{abstract}


\IEEEpeerreviewmaketitle

\section{Introduction}
\IEEEPARstart{R}{ecently}, reconfigurable intelligent surfaces (RISs) and their facilitated smart radio environment have attracted significant interest from both industry and academia in order to enhance the performance of beyond $5$G and $6$G communications \cite{j:ZhouPoor2024,j:MuDhahir2024,j:YangXuRenzo2024}. Nevertheless, to date, most of the relevant state-of-the-art research focuses on the modeling and/or optimization of RIS-enabled communications based on the assumption that a single-only network operator controls and/or utilizes RIS(s). Such a system model assumption seems superficial when considering a multi-operator/multi-RIS environment. In reality, different wireless network operators coexist \emph{independently} within a given geographical area, each functioning using its own RIS(s). Consequently, multiple users simultaneously receive a transmission impact from different operators in their (potentially close) vicinity. In fact, the case when a RIS deployment optimized by one operator serving the needs of its subscribed users causes an unavoidable impact on the performance of users served by other operators. This effect, entitled as inter-operator interference (IOI), has not been studied in the appropriate depth so far. 

More specifically, quite a few research studies analyze the system performance under a multi-operator/multi-RIS environment \cite{j:CaiRang2022,j:Schwarz2024,j:Gurgunoglu2023,j:yashvanth2023}. In \cite{j:CaiRang2022} and \cite{j:Schwarz2024}, the RIS phase shift configuration is jointly optimized in the presence of multiple operators; however, requiring inter-operator coordination which is not always feasible or desirable. In \cite{j:Gurgunoglu2023}, the important problem of pilot contamination during channel-state information (CSI) acquisition in a multi-operator/multi-RIS environment is studied. Most recently, \cite{j:yashvanth2023} analyzed the performance of a multi-operator communication system; yet, assuming that only a single operator is equipped with a corresponding RIS technology. All these studies assumed either a multi-operator/single-RIS infrastructure and/or Rayleigh-only channel fading. Quite recently, \cite{j:YashvanthMurthy2024} studied the case when multiple RISs from the same operator are used to serve users in the presence of another operator communicating to its own system users without utilizing RIS. It turns out from \cite{j:yashvanth2023,j:YashvanthMurthy2024} that system subscribers of the operator that does not utilize RIS technology benefit from the presence of RIS(s), even if they do not explicitly operate on their behalf. In our recent work \cite{j:MiridakisPopovski2024}, we analyzed the impact of IOI in a multi-operator multi-RIS environment (where each operator controls its own RIS), yet assuming independent-only channel fading for the RIS-related links as well as a time-synchronized IOI, which does not entirely reflect on practical communication conditions. To our knowledge, the performance analysis of a multi-operator/multi-RIS communication system under the presence of spatially correlated RISs and/or random multiple IOI sources (which is practical) is not available in the open technical literature so far.  

Building upon these observations, we analytically study the impact of IOI in the realistic condition of a multi-operator multi-RIS networking system, under the presence of spatially correlated RIS arrays of arbitrary size. Each RIS-enabled operator may introduce IOI, individually, which is initiated and/or terminated at random time instances within the transmission frame whereon the reference system operates (since different operators do not control the transmission of one another). Regarding some key performance indicators, tight lower and upper bounds as well as asymptotic (in the large RIS array regime) closed-form expressions of the system spectral efficiency are obtained. The derived results are valid for the versatile correlated Nakagami-$m$ channel fading model. Prominent engineering insights are revealed, such as the impact of uncontrolled (multiple) RIS(s) on the total communication quality and the influence of spatial correlation on the system performance.

{\color{black}The contributions and key insights of this work are summarized as follows:
\begin{itemize}
	\item For correlated Nakagami-$m$ faded channels and an asynchronous multi-RIS environment with uncontrolled IOI, new and tight closed-form lower and upper bounds for the system spectral efficiency are obtained.
	\item In the asymptotically high RIS-array regime, IOI tends to vanish. Yet, the spatial correlation at RIS still affects the system performance.
	\item Yet, the latter spatial correlation at RIS does not dramatically influence the system performance, even when there is no line-of-sight signal propagation.
	\item The presence of multiple (uncontrolled, asynchronous) IOI distinct sources (i.e., originated from different operators) seems to act beneficially to the overall system performance.
\end{itemize}
}

{\textbf{Notation}:} Vectors and matrices are represented by lowercase and uppercase bold typeface letters, respectively. A diagonal matrix with entries $x_{1},\cdots,x_{n}$ is defined as $\diag\{x_{i}\}^{n}_{i=1}$. Superscript $(\cdot)^{T}$ denotes transpose; $|\cdot|$ represents absolute value, $\angle[\cdot]$ is the phase of a complex argument, and ${\rm j}\triangleq \sqrt{-1}$. $\mathbb{E}[\cdot]$ is the expectation operator, $\mathbb{E}_{X}[\cdot]$ denotes expectation with respect to the random variable (RV) $X$, symbol $\overset{\text{d}}=$ means equality in distribution and $\overset{\text{d}}\approx$ defines almost sure convergence (asymptotically) in distribution. $f_{X}(\cdot)$ and $F_{X}(\cdot)$ represent the probability density function (PDF) and cumulative distribution function (CDF) of $X$, respectively. $\mathcal{CN}(\mu,v)$ defines a complex-valued Gaussian RV with mean $\mu$ and variance $v$. Also, $\Gamma(\cdot)$ denotes the Gamma function \cite[Eq. (8.310.1)]{tables}; $\psi(\cdot)$ is the Digamma function \cite[Eq. (8.360.1)]{tables}; $I_{0}(\cdot)$ is the zeroth order modified Bessel function of the first kind \cite[Eq. (8.445)]{tables}; $Q_{1}(\cdot,\cdot)$ is the first-order Marcum $Q$-function; ${}_2F_{1}\left(\cdot,\cdot;\cdot;\cdot\right)$ is the Gauss hypergeometric function \cite[Eq. (9.100)]{tables}; and $G[\cdot|\cdot]$ represents the Meijer's $G-$function \cite[Eq. (9.301)]{tables}.

\section{System and Signal Model}
\begin{figure}[!t]
\centering
\includegraphics[trim=.5cm .5cm .5cm 0.0cm, clip=true,totalheight=0.31\textheight]{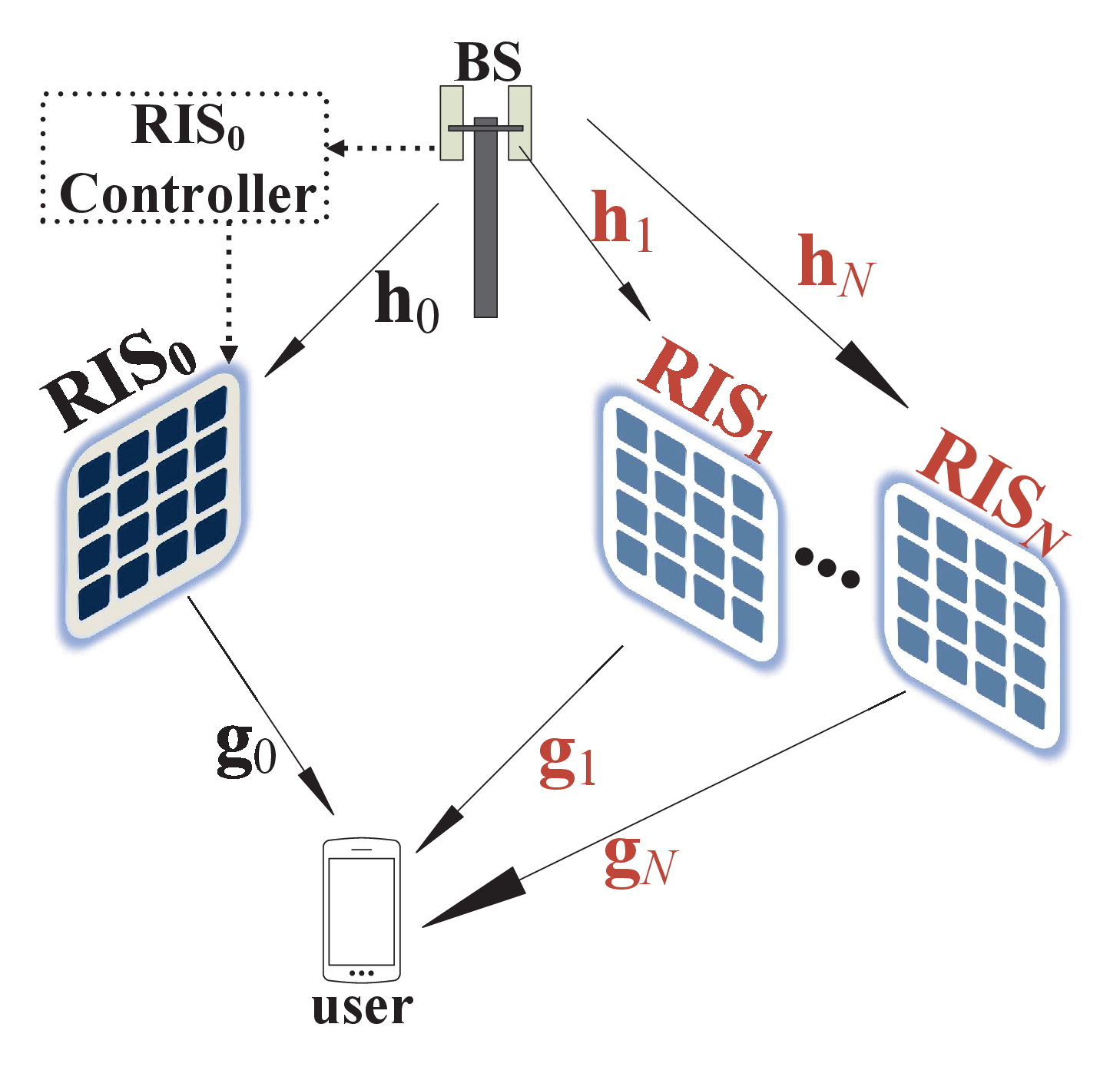}
\caption{The considered system model is sketched, where a base station (BS) stands for the transmitter, the receiver is the reference system user and the communication is aided through ${\rm RIS}_{0}$ which is directly controlled by its corresponding operator. Meanwhile, $\{{\rm RIS}_{i}\}^{N}_{i=1}$ operate \emph{independently} in a close vicinity, which are controlled by different (distinct) operators.}
\label{fig1}
\end{figure}

Consider a wireless communication system with a single-antenna transmitter and receiver operating over a quasi-static block-fading channel. The end-to-end communication is assisted by an intermediate RIS equipped with an array of arbitrary-size passive elements. According to \cite{j:EmilLuca2021}, channel fading (spatial) correlation is always present in practical RIS illustrations due to the rectangular structure of RIS arrays. Thereby, it is assumed that the transmitter-to-RIS and RIS-to-receiver links undergo correlated Nakagami-$m$ channel fading due to their relatively close distance and the (moderate) presence of a line-of-sight (LoS) channel gain component. In addition, other operators coexist in a close vicinity of the considered communication system, which likewise utilize RIS-enabled communication for their own subscribers; viz. Fig.~\ref{fig1}. 

Hereinafter, we term ${\rm RIS}_{0}$ the RIS controlled by the reference operator, equipped with $M_{0}$ elements, and ${\rm RIS}_{i}$ with $1\leq i \leq N$ another RIS controlled by a different operator, equipped with $M_{i}$ elements. Although the multiple operators in principle occupy different spectrum bands and there is no out-of-band interference, the aforementioned infrastructure includes RISs with different features whereby a new type of interference inherently emerges; entitled as IOI \cite{j:MiridakisPopovski2024}. It is caused by rapidly sudden changes in the phase shifts of the uncontrolled RIS (say, ${\rm RIS}_{1}$ of Fig.~\ref{fig1}), which unavoidably result to an incoherent sum of the received channel gain at the reference user. Due to the possibly different array sizes and mutually independent (uncontrolled) phase shifts of ${\rm RIS}_{0}$ and ${\rm RIS}_{i}\:\forall i$, we hereafter term these RISs as \emph{heterogeneous}.   

\begin{figure}[!t]
\centering
\includegraphics[trim=.0cm .0cm .0cm 0.0cm, clip=true,totalheight=0.22\textheight]{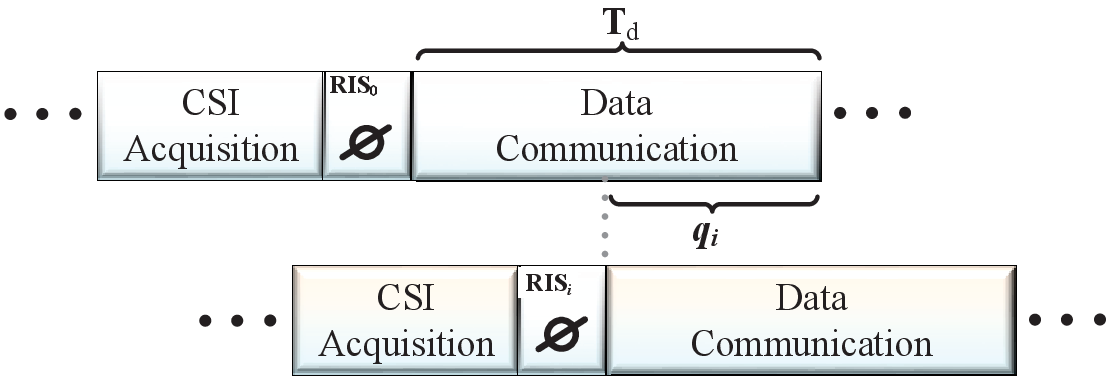}
\caption{A typical example of uncontrolled IOI due to the independent phase shift adjustment at the external $i^{\rm th}$ RIS. Symbol $\varnothing$ denotes the phase updating process at each RIS.}
\label{fig2}
\end{figure}

More specifically, the received signal reads as\footnote{Typically, the channel gain from the direct (transmitter-to-receiver) link is available at the receiver side. However, we assume that its gain is almost negligible due to, e.g., a high propagation loss and/or long link-distance. Further, our prime goal is to emphasize on the impact of spatially correlated RIS-enabled links.}
\begin{align}
{\rm r}=\sqrt{\rm p}\left(\mathbf{h}^{T}_{0}\mathbf{\Phi}_{0}\mathbf{g}_{0}+\sum^{N}_{i=1}\sqrt{q_{i}}\mathbf{h}^{T}_{i}\mathbf{\Phi}_{i}\mathbf{g}_{i}\right)s+w,
\label{received}
\end{align}
where ${\rm p}$ is the transmit signal-to-noise ratio (SNR) plus other fixed or predetermined path losses (e.g., path-loss signal propagation attenuation, antenna gains, noise figure at the receiver, and large-scale channel fading), $s\in \mathbb{C}$ is the unit-power transmit signal, $\mathbf{h}_{0}\in \mathbb{C}^{M_{0} \times 1}$ and $\mathbf{g}_{0}\in \mathbb{C}^{M_{0} \times 1}$ are the channel vectors of the transmitter-to-${\rm RIS}_{0}$ and ${\rm RIS}_{0}$-to-receiver links, respectively; $\mathbf{\Phi}_{0}\triangleq \diag\{e^{{\rm j} \phi_{i}}\}^{M_{0}}_{i=1}$ denotes the phase rotations at ${\rm RIS}_{0}$; and $w\in \mathbb{C}$ defines the additive white Gaussian noise at the receiver such that $w\overset{\text{d}}=\mathcal{CN}(0,1)$. In the same analogy, $\mathbf{h}_{i}\in \mathbb{C}^{M_{i} \times 1}$, $\mathbf{g}_{i}\in \mathbb{C}^{M_{i} \times 1}$ and $\mathbf{\Phi}_{i}\in \mathbb{C}^{M_{i} \times M_{i}}$ are formulated with $1\leq i\leq N$. Due to the heterogeneous nature of ${\rm RIS}_{i}$, $q_{i}$ in \eqref{received} denotes an uncertainty factor with respect to the amount of impact (or presence) of $\mathbf{h}^{T}_{i}\mathbf{\Phi}_{i}\mathbf{g}_{i}$ at the received signal of the reference system. In particular, $q_{i}$ denotes the percentage of IOI (caused by ${\rm RIS}_{i}$) within the data transmission interval of the reference system, as illustrated in Fig.~\ref{fig2}. Notably, it cannot be controlled by the reference system; and by doing so, it is reasonable to be modeled as a uniformly distributed RV (i.e., equiprobable) in the interval $[0,{\rm T}_{\rm d}]$, where ${\rm T}_{\rm d}$ stands for the data transmission time of the reference system (c.f., Fig.~\ref{fig2}).\footnote{{\color{black}Hereinafter, it is assumed that ${\rm T}_{\rm d}\leq {\rm T}_{{\rm d},i}$, where ${\rm T}_{{\rm d},i}$ denotes the data transmission time of the $i^{\rm th}$ operator. It is shown in Appendix that the derived results serve as lower performance bounds for the opposite case when ${\rm T}_{\rm d}>{\rm T}_{{\rm d},i}$.}} 

Channel vectors $\mathbf{h}_{0}$ and $\mathbf{g}_{0}$ follow the correlated Nakagami-$m$ distribution each having a spatial correlation matrix $\mathbf{C}_{\mathbf{u}}$ with $\mathbf{u}\in \{\mathbf{h}_{0},\mathbf{g}_{0}\}$, while its entry at the $i^{\rm th}$ row and $j^{\rm th}$ column is given by
\begin{align}
\mathbf{c}^{\mathbf{u}}_{i,j}=\rho^{\left\|\mathbf{c}_{i}-\mathbf{c}_{j}\right\|/c_{0}},
\end{align}
where $\{i,j\}\in[1,M_{0}]$; $\mathbf{c}_{i}\triangleq [0,l y(i),k z(i)+l_{0}]$; $y(i)\triangleq {\rm mod}(i-1,M_{0,h})$; $z(i)\triangleq \lfloor \frac{i-1}{M_{0,h}}\rfloor$; $l$ and $k$ denote the horizontal and vertical lengths of each RIS element; $l_{0}$ is the RIS height from the origin; $M_{0,h}$ is the number of RIS elements in the horizontal direction (such that $M_{0}\triangleq M_{0,h}M_{0,v}$ with $M_{0,v}$ standing for the corresponding number at the vertical direction of RIS array); $c_{0}$ is the smallest element spacing at RIS; and $\rho \in [0,1]$ is the power correlation coefficient of RIS array \cite{j:WangCoon2024}. Also, each entry of $\mathbf{h}_{0}$ and $\mathbf{g}_{0}$ has a Nakagami shape parameter $m_{\mathbf{h}}$ and $m_{\mathbf{g}}$, respectively, as well as it is modeled by a unit scale. 

By applying coherent detection, i.e., given $\{\mathbf{h}_{0},\mathbf{g}_{0}\}$, the reference operator strives to maximize the received RIS-enabled gains so as to obtain the maximum achievable SNR. Then, according to \cite[Eq. (28)]{j:WuZhang2019}, the optimum setup of the $l^{\rm th}$ phase shift element of ${\rm RIS}_{0}$ becomes $\phi_{l}=e^{{\rm j} (-\angle[\mathbf{h}_{0,l}]-\angle[\mathbf{g}_{0,l}])}$ where $l\in [1,M_{0}]$. Hence, the post-detected signal at the receiver stems as \cite{j:Badiu2020}  
\begin{align}
\nonumber
{\rm y}&=e^{-{\rm j} \angle[\mathbf{h}^{T}_{0}\mathbf{\Phi}_{0}\mathbf{g}_{0}]}{\rm r}\\
\nonumber
&=\sqrt{\rm p}\left|\mathbf{h}^{T}_{0}\mathbf{g}_{0}\right| e^{{\rm j}\theta} s\\
&\ \ \ +e^{-{\rm j} \angle[\mathbf{h}^{T}_{0}\mathbf{\Phi}_{0}\mathbf{g}_{0}]} \left(\sum^{N}_{i=1}\sqrt{{\rm p} q_{i}}\mathbf{h}^{T}_{i}\mathbf{\Phi}_{i}\mathbf{g}_{i} s+w\right),
\label{detected}
\end{align}
where $\theta$ is the residual phase mismatch caused by imperfect (yet, practical \cite{j:MiridakisTsifYao2023}) CSI acquisition at the receiver regarding the channel gains of the reference system links $\mathbf{h}_{0}$ and $\mathbf{g}_{0}$. On the other hand, it does not control nor is aware of the CSI-related links regarding ${\rm RIS}_{i}$. Typically, $\theta$ can be modeled as a zero-mean von-Mises RV whose concentration parameter, defined as $\kappa$, captures the accuracy of channel estimation (a lower $\kappa$ stands for an increased estimation error). Also, $e^{-{\rm j} \angle[\mathbf{h}^{T}_{0}\mathbf{\Phi}_{0}\mathbf{g}_{0}]}w\overset{\text{d}}=w$ due to the isotropic identity of circularly symmetric Gaussian RVs. The resultant received SNR at the reference user yields as
\begin{align}
\gamma\triangleq &{\rm p} \bigg| \left|\mathbf{h}^{T}_{0}\mathbf{g}_{0}\right| e^{{\rm j}\theta}+e^{-{\rm j} \angle[\mathbf{h}^{T}_{0}\mathbf{\Phi}_{0}\mathbf{g}_{0}]} \sum^{N}_{i=1}\sqrt{q_{i}}\mathbf{h}^{T}_{i}\mathbf{\Phi}_{i}\mathbf{g}_{i}\bigg|^{2},
\label{snrrr}
\end{align}
where the left-most sum term of \eqref{snrrr} is considered as the (seemingly) known part of channel gain, while the right-most term is the (unknown) source of randomness due to the presence of ${\rm RIS}_{i}$, which in turn causes a potentially strong fluctuation of the above SNR.

\section{System Performance}
For ease of presentation, let $\mathcal{X}\triangleq \left|\mathbf{h}^{T}_{0}\mathbf{g}_{0}\right| e^{{\rm j}\theta}$ and $\mathcal{Y}\triangleq e^{-{\rm j} \angle[\mathbf{h}^{T}_{0}\mathbf{\Phi}_{0}\mathbf{g}_{0}]} \sum^{N}_{i=1}\sqrt{q_{i}}\mathbf{h}^{T}_{i}\mathbf{\Phi}_{i}\mathbf{g}_{i}$. Then, the received SNR can be compactly rewritten as $\gamma={\rm p}|\mathcal{X}+\mathcal{Y}|^{2}$. To analyze the performance of $\gamma$, we proceed by computing the individual statistics of $\mathcal{X}$ and $\mathcal{Y}$. To this end, it holds that
\begin{align}
\nonumber
\mathbb{E}[\mathcal{X}]&=\sum^{M_{0}}_{i=1}\mathbb{E}[\mathbf{h}_{0,i}] \mathbb{E}[\mathbf{g}_{0,i}] \mathbb{E}[e^{{\rm j}\theta}]\\
&=\frac{\Gamma(m_{\mathbf{h}}+\frac{1}{2})\Gamma(m_{\mathbf{g}}+\frac{1}{2})}{\Gamma(m_{\mathbf{h}})\Gamma(m_{\mathbf{g}})\sqrt{m_{\mathbf{h}} m_{\mathbf{g}}}}M_{0} \overline{\theta},
\label{expX}
\end{align}
where $\overline{\theta}\triangleq \mathbb{E}[e^{{\rm j}\theta}]=\frac{I_{1}(\kappa)}{I_{0}(\kappa)}$ or $\overline{\theta}=1$ for a perfect CSI acquisition. Further, the second moment of $\mathcal{X}$ reads as
\begin{align}
\nonumber
\mathbb{E}[\mathcal{X}^{2}]&=\mathbb{E}\left[\sum^{M_{0}}_{i=1}\mathbf{h}^{2}_{0,i} \mathbf{g}^{2}_{0,i}\right]\\
\nonumber
&\ \ +2 \mathbb{E}\left[\sum^{M_{0}-1}_{i=1}\sum^{M_{0}}_{k=i+1}\mathbf{h}_{0,i} \mathbf{g}_{0,i}{\rm cos}(\phi_{i})\mathbf{h}_{0,k} \mathbf{g}_{0,k}{\rm cos}(\phi_{k})\right]\\
\nonumber
&\ \ +2 \mathbb{E}\left[\sum^{M_{0}-1}_{i=1}\sum^{M_{0}}_{k=i+1}\mathbf{h}_{0,i} \mathbf{g}_{0,i}{\rm sin}(\phi_{i})\mathbf{h}_{0,k} \mathbf{g}_{0,k}{\rm sin}(\phi_{k})\right]\\
&=M_{0} (1-\overline{\theta}^{2})+{\rm Tr}[\mathbf{R}_{\mathbf{h}}\mathbf{R}_{\mathbf{g}}]\overline{\theta}^{2},
\label{expX2}
\end{align}
where the last expression arises by utilizing the linear property of expectation operation; using that $\mathbb{E}[{\rm cos}(\phi_{i})]=\overline{\theta}$ while amplitude and phase coefficients are mutually independent RVs; and the fact that sine is an odd function (thus, the third term of the first equation becomes zero). Also, the second term of the first equation evaluates to $\overline{\theta}^{2}(\mathbb{E}[\mathbf{h} \mathbf{h}^{T}]\mathbb{E}[\mathbf{g} \mathbf{g}^{T}]-M_{0})$. Finally, $\mathbf{R}_{\mathbf{u}}\triangleq \mathbb{E}[\mathbf{u} \mathbf{u}^{T}]$ is defined in \eqref{expX2} with $\mathbf{u}\in\{\mathbf{h},\mathbf{g}\}$ and $[\mathbf{R}_{\mathbf{u}}]_{i,j}$ with $\{i,j\}\in [1,M_{0}]$ is computed by \cite[Eq. (25)]{j:WangCoon2024}
\begin{align}
\left[\mathbf{R}_{\mathbf{u}}\right]_{i,j}=\frac{\Gamma^{2}(m_{\mathbf{u}}+\frac{1}{2})}{\Gamma^{2}(m_{\mathbf{u}}) m_{\mathbf{u}}}{}_2F_1\left(-\frac{1}{2},-\frac{1}{2};m_{\mathbf{u}};\mathbf{c}^{\mathbf{u}}_{i,j}\right).
\label{Ru}
\end{align} 
For the special (ideal) case of independent channel fading at RIS, it holds that $\mathbf{c}^{\mathbf{u}}_{i,j}=0$ for $i\neq j$ while $\mathbf{c}^{\mathbf{u}}_{i,i}=1$. In addition, the variance of $\mathcal{X}$ is directly obtained by
\begin{align}
\nonumber
{\rm Var}[\mathcal{X}]&=\mathbb{E}[\mathcal{X}^{2}]-\mathbb{E}^{2}[\mathcal{X}]\\
\nonumber
&=M_{0} (1-\overline{\theta}^{2})+{\rm Tr}[\mathbf{R}_{\mathbf{h}}\mathbf{R}_{\mathbf{g}}]\overline{\theta}^{2}\\
&\ \ -\left(\frac{\Gamma(m_{\mathbf{h}}+\frac{1}{2})\Gamma(m_{\mathbf{g}}+\frac{1}{2})}{\Gamma(m_{\mathbf{h}})\Gamma(m_{\mathbf{g}})\sqrt{m_{\mathbf{h}} m_{\mathbf{g}}}}M_{0} \overline{\theta}\right)^{2}.
\label{varX}
\end{align}
Via the moment-matching method, the PDF/CDF of $\mathcal{X}$ can be efficiently approached by a Gamma distribution as
\begin{align}
f_{\mathcal{X}}(x)\approx \frac{x^{\alpha-1}\exp\left(-\frac{x}{\beta}\right)}{\Gamma(\alpha)\beta^{\alpha}}\quad \textrm{and}\quad F_{\mathcal{X}}(x)\approx 1-\frac{\Gamma(\alpha,\frac{x}{\beta})}{\Gamma(\alpha)},
\label{pdfcdfX}
\end{align}
where
\begin{align}
\alpha \triangleq \mathbb{E}^{2}[\mathcal{X}]/{\rm Var}[\mathcal{X}]\quad \textrm{and}\quad \beta \triangleq {\rm Var}[\mathcal{X}]/\mathbb{E}[\mathcal{X}].
\label{params}
\end{align}

Regarding the distribution of $\mathcal{Y}$, notice that for a given $q_{i}$ and fixed $\mathbf{h}^{T}_{0}\mathbf{\Phi}_{0}\mathbf{g}_{0}$, the resultant phase of $\sqrt{q_{i}}\mathbf{h}^{T}_{i}\mathbf{\Phi}_{i}\mathbf{g}_{i} e^{-{\rm j} \angle[\mathbf{h}^{T}_{0}\mathbf{\Phi}_{0}\mathbf{g}_{0}]}$ is uniformly distributed in $[-\pi,\pi)$ since the reference system knows nothing about $\Phi_{i}\:\forall i\in [1,N]$. Thereby, let $\mathcal{Y}_{i}\triangleq \sqrt{q_{i}}\mathbf{h}^{T}_{i}\mathbf{\Phi}_{i}\mathbf{g}_{i} e^{-{\rm j} \angle[\mathbf{h}^{T}_{0}\mathbf{\Phi}_{0}\mathbf{g}_{0}]}$. It holds that $\mathbb{E}[\mathcal{Y}_{i}]=0$ due to the uniform phase. Also, ${\rm Var}[\mathcal{Y}_{i}]=\mathbb{E}[q_{i}] \mathbb{E}[|\mathbf{h}^{T}_{i}\mathbf{\Phi}_{i}\mathbf{g}_{i} e^{-{\rm j} \angle[\mathbf{h}^{T}_{0}\mathbf{\Phi}_{0}\mathbf{g}_{0}]}|^{2}]$ because $\mathbb{E}[\mathcal{Y}_{i}]=0$. Since $q_{i}$ is uniformly distributed in\footnote{In fact, $q_{i}$ is uniformly distributed in $[0,1]\times {\rm T}_{\rm d}$. However, for ease of clarity and without loss of generality, the normalized version of ${\rm T}_{\rm d}$ is used herein, i.e., ${\rm T}_{\rm d}=1$.} $[0,1]$, we get $\mathbb{E}[q_{i}]=1/2$. Moreover, by expanding $|\mathbf{h}^{T}_{i}\mathbf{\Phi}_{i}\mathbf{g}_{i} e^{-{\rm j} \angle[\mathbf{h}^{T}_{0}\mathbf{\Phi}_{0}\mathbf{g}_{0}]}|^{2}$ in quite a similar basis as per the first equality of \eqref{expX2} while noticing that the underlying phase of $\mathcal{Y}_{i}$ is uniform, we arrive at $\mathbb{E}[|\mathbf{h}^{T}_{i}\mathbf{\Phi}_{i}\mathbf{g}_{i} e^{-{\rm j} \angle[\mathbf{h}^{T}_{0}\mathbf{\Phi}_{0}\mathbf{g}_{0}]}|^{2}]=M_{i}$. Finally, it stems that ${\rm Var}[\mathcal{Y}_{i}]=M_{i}/2$ and thus the total variance of $\mathcal{Y}$ becomes ${\rm Var}[\mathcal{Y}]=\sum^{N}_{i=1}M_{i}/2$ (because $\{\mathcal{Y}_{i}\}^{N}_{i=1}$ are mutually independent RVs).

Then, the Lyapunov central limit theorem can be invoked, such as $\mathcal{Y}_{i}\overset{\text{d}}\approx \mathcal{CN}(0,M_{i}/2),\: M_{i}\rightarrow +\infty$ and 
\begin{align}
\mathcal{Y}\overset{\text{d}}\approx \mathcal{CN}\left(0,\sum^{N}_{i=1}M_{i}/2\right).
\label{distrY}
\end{align}
Moreover, conditioned on the estimated channel gain $\mathcal{X}$, the total received channel gain $\mathcal{X}+\mathcal{Y}\overset{\text{d}}\approx \mathcal{CN}(\mathcal{X},\sum^{N}_{i=1}M_{i}/2)$. It turns out that the CDF of the resultant (conditional) SNR is expressed as
\begin{align}
F_{\gamma|\mathcal{X}}(\gamma|\mathcal{X})\approx 1-Q_{1}\left(\frac{\mathcal{X}}{\sqrt{\sigma^{2}}},\sqrt{\frac{\gamma}{p \sigma^{2}}}\right),
\label{CDFSNRcon}
\end{align}
where $\sigma^{2}\triangleq \sum^{N}_{i=1}M_{i}/4$ is defined for notational clarity. It follows that the corresponding unconditional CDF of the received SNR can be approached by
\begin{align}
\nonumber
F_{\gamma}(\gamma)&\approx 1-\int^{\infty}_{0}Q_{1}\left(\frac{x}{\sqrt{\sigma^{2}}},\sqrt{\frac{\gamma}{p \sigma^{2}}}\right)\frac{x^{\alpha-1}\exp\left(-\frac{x}{\beta}\right)}{\Gamma(\alpha)\beta^{\alpha}}{\rm d}x\\
&=1-\beta \int^{1}_{0}Q_{1}\left(\frac{\beta {\rm ln}\left(\frac{1}{x}\right)}{\sqrt{\sigma^{2}}},\sqrt{\frac{\gamma}{p \sigma^{2}}}\right)\frac{\left[\beta {\rm ln}\left(\frac{1}{x}\right)\right]^{\alpha-1}}{\Gamma(\alpha)\beta^{\alpha}}{\rm d}x,
\label{CDFSNRuncon}
\end{align}
where the latter equality arises by utilizing integration by substitution (i.e., $e^{-x/\beta}\leftarrow x$) and preserves numerical stability compared to the former one. Outage performance of the considered system is directly obtained via \eqref{CDFSNRuncon} by simply interchanging the auxiliary constant $\gamma$ with $\gamma_{\rm th}$ where $\gamma_{\rm th}$ denotes a predetermined SNR outage threshold.

We proceed by further analyzing the system spectral efficiency, which is defined as $\overline{C}\triangleq \mathbb{E}[{\rm log}_{2}(1+\gamma)]$. Unfortunately, a closed-form solution of $\overline{C}$ is not feasible mainly due to the underlying distribution of $\gamma$ which is quite involved. However, the upper and lower bounds of the system spectral efficiency admit closed-form expressions, which are presented hereinafter.

Regarding the upper bound of $\overline{C}$, it holds that
\begin{align}
\nonumber
\overline{C}&\leq {\rm log}_{2}\left(1+{\rm p} \mathbb{E}\left[|\mathcal{X}+\mathcal{Y}|^{2}\right]\right)\\
&={\rm log}_{2}\left(1+{\rm p} [\alpha (\alpha+1) \beta^{2}+2 \sigma^{2}]\right),
\label{cupper}
\end{align}  
where the Jensen's inequality and the log-concavity was used for the derivation of the latter expression as well as the fact that
\begin{align}
\nonumber
\mathbb{E}\left[|\mathcal{X}+\mathcal{Y}|^{2}\right]&=\int^{\infty}_{0}\mathbb{E}_{\mathcal{X}}\left[|\mathcal{X}+\mathcal{Y}|^{2}|\mathcal{X}\right]f_{\mathcal{X}}(x){\rm d}x\\
&=\int^{\infty}_{0}\left(x^{2}+2 \sigma^{2}\right) f_{\mathcal{X}}(x){\rm d}x.
\label{cupperdef}
\end{align} 
The lower bound of $\overline{C}$ reads as
\begin{align}
\overline{C}\geq {\rm log}_{2}\left(1+{\rm p} \exp\left(\mathbb{E}\left[{\rm ln}\left(|\mathcal{X}+\mathcal{Y}|^{2}\right)\right]\right)\right),
\label{clower}
\end{align} 
since ${\rm log}_{2}(1+\exp(z))$ is convex on $z$. Thereby, the computation of $\mathbb{E}[{\rm ln}(|\mathcal{X}+\mathcal{Y}|^{2})]$ is required. Recall that $|\mathcal{X}+\mathcal{Y}|^{2}$, conditioned on $\mathcal{X}$, is typically a complex-valued non-central chi-squared RV and a corresponding PDF given by
\begin{align}
f_{|\mathcal{X}+\mathcal{Y}|^{2}|\mathcal{X}}(z)=\frac{\exp\left(-\frac{(\mathcal{X}^{2}+z)}{2 \sigma^{2}}\right)}{2 \sigma^{2}} I_{0}\left(\frac{\mathcal{X} \sqrt{z}}{\sigma^{2}}\right).
\label{pdfX2}
\end{align}
Then, utilizing \cite[Eq.~(13)]{c:moser2008} and after performing some straightforward manipulations, we get
\begin{align}
\mathbb{E}\left[{\rm ln}\left(|\mathcal{X}+\mathcal{Y}|^{2}\right)\right]=\mathbb{E}_{\mathcal{X}}\left[{\rm ln}\left(\mathcal{X}^{2}\right)+\Gamma\left(\frac{\mathcal{X}^{2}}{2 \sigma^{2}}\right)\right].
\label{condclower}
\end{align}
The desired result yields as
\begin{align}
\nonumber
&\mathbb{E}\left[{\rm ln}\left(|\mathcal{X}+\mathcal{Y}|^{2}\right)\right]=\int^{\infty}_{0}\left[{\rm ln}\left(x^{2}\right)+\Gamma\left(\frac{x^{2}}{2 \sigma^{2}}\right)\right] f_{\mathcal{X}}(x){\rm d}x\\
&=2 [\psi(\alpha)+{\rm ln}(\beta)]+\frac{2^{\alpha-1}}{\Gamma(\alpha)\sqrt{\pi}} G^{2,2}_{3,2}\left[\frac{2 \beta^{2}}{\sigma^{2}}~\vline
\begin{array}{c}
\frac{1-\alpha}{2},1-\frac{\alpha}{2},1 \\
0,0 
\end{array} \right],
\label{uncondclower}
\end{align}
where the latter closed-form expression is obtained with the aid of \cite[Eq. (4.352.1)]{tables} and \cite[Eqs. (8.416.2) and (2.24.3.1)]{b:prudnikovvol3}, while using \eqref{pdfcdfX}. Finally, the spectral efficiency is lower bounded by putting \eqref{uncondclower} into \eqref{clower}.

To gain more insights, we further resort to an asymptotic analysis when the RIS arrays include a relatively high number of passive elements, i.e., when $\{M_{0},M_{i}\}^{N}_{i=1}\rightarrow +\infty$. Recall from \eqref{snrrr} that $\mathcal{Y}\triangleq e^{-{\rm j} \angle[\mathbf{h}^{T}_{0}\mathbf{\Phi}_{0}\mathbf{g}_{0}]} \sum^{N}_{i=1}\sqrt{q_{i}}\mathbf{h}^{T}_{i}\mathbf{\Phi}_{i}\mathbf{g}_{i}$. Further, as $M_{i}\rightarrow +\infty$, the channel hardening condition appears such that $\mathcal{Y}\rightarrow \mathbb{E}[\mathcal{Y}]=0$ since $e^{-{\rm j} \angle[\mathbf{h}^{T}_{0}\mathbf{\Phi}_{0}\mathbf{g}_{0}]}\mathbf{\Phi}_{i}$ averages to zero $\forall i$ because of its underlying uniform distribution (also see \cite[Thm.~3.7]{b:CoullietDebbah2011}). Hence, $\gamma \rightarrow {\rm p} \mathcal{X}^{2}$.

With the aid of \eqref{expX} and \eqref{varX}, we can easily show that $\mathbb{E}[\mathcal{X}]/\sqrt{{\rm Var}[\mathcal{X}]}\propto \sqrt{M_{0}}$ for a growing $M_{0}$, thus channel hardening applies in a similar basis as in \cite[Eq. (19)]{j:IbrahimTabassum2021}, yielding $\mathcal{X}^{2}\rightarrow \mathbb{E}[\mathcal{X}^{2}]=M_{0} (1-\overline{\theta}^{2})+{\rm Tr}[\mathbf{R}_{\mathbf{h}}\mathbf{R}_{\mathbf{g}}]\overline{\theta}^{2}$ as per \eqref{expX2}, and finally, we arrive at
\begin{align}
\gamma \rightarrow {\rm p} [M_{0} (1-\overline{\theta}^{2})+{\rm Tr}[\mathbf{R}_{\mathbf{h}}\mathbf{R}_{\mathbf{g}}]\overline{\theta}^{2}]< {\rm p} M_{0},
\label{asysnr}
\end{align}
which holds for relatively large arrays and indicates the impact of the underlying spatial correlation at RIS. {\color{black}It is also noteworthy from \eqref{asysnr} that the effect of IOI tends to vanish at the asymptotically high RIS-array regime. Nonetheless, the received SNR scales linearly with the number of RIS elements (and not quadratically as in classical systems without IOI). This linear scaling has also been observed in some previous studies \cite{j:MiridakisPopovski2024,j:YashvanthMurthy2024}.}

\section{Numerical Results and Outcomes}
In this section, the derived analytical results are verified via numerical validation (in line-curves), whereas they are cross-compared with corresponding Monte-Carlo simulations (in solid circle mark-signs). Without loss of generality and for ease of clarity, we set the power correlation coefficient at RIS arrays as $\rho=1/2$ (unless otherwise specified) and an identical Nakagami shape parameter per hop, i.e., $m_{\mathbf{h}}=m_{\mathbf{g}}\triangleq m$. {\color{black}For convenience, the main simulation parameters are given in Table~\ref{table}.}

\begin{table}[!t]
\caption {Simulation Parameters} 
\begin{center}
\begin{tabular}{l| l}\hline
\textbf{Parameter} & \textbf{Symbol}\\\hline
\# of RIS elements at the reference operator & $M_{0}$ \\
\# of RIS elements at the $i^{\rm th}$ operator  & $M_{i}$ \\
\# of operators (except of the reference one) & $N$  \\
Concentration parameter of phase mismatch $\theta$ & $\kappa$\\
Nakagami shape parameter & $m$\\
Power correlation coefficient of RIS array & $\rho$ \\
Transmit SNR & ${\rm p}$ \\\hline
\end{tabular}
\end{center}
\label{table}
\end{table}

\begin{figure}[!t]
\centering
\includegraphics[trim=2.5cm 0.0cm 2.5cm 1.0cm, clip=true,totalheight=0.35\textheight]{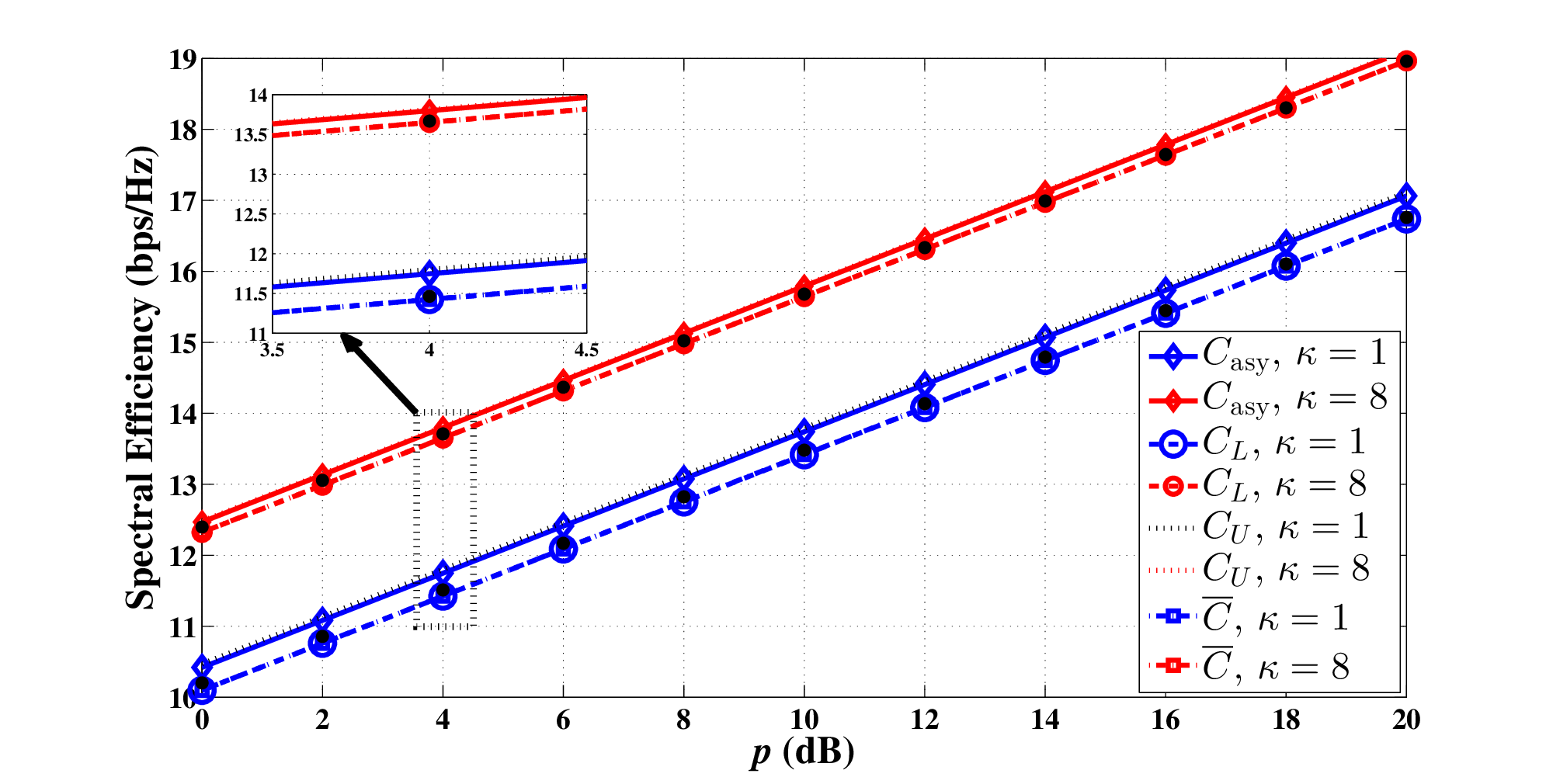}
\caption{Spectral efficiency vs. various values of the transmit SNR, where $N=1$ and $\{M_{0},M_{1}\}=100$ (i.e., two operators each equipped with a $10\times 10$ RIS array) and $\{m_{\mathbf{h}},m_{\mathbf{g}}\}=1$ (i.e., Rayleigh channel fading).}
\label{fig3}
\end{figure}

\begin{figure}[!t]
\centering
\includegraphics[trim=2.5cm 0.0cm 2.5cm 1.0cm, clip=true,totalheight=0.35\textheight]{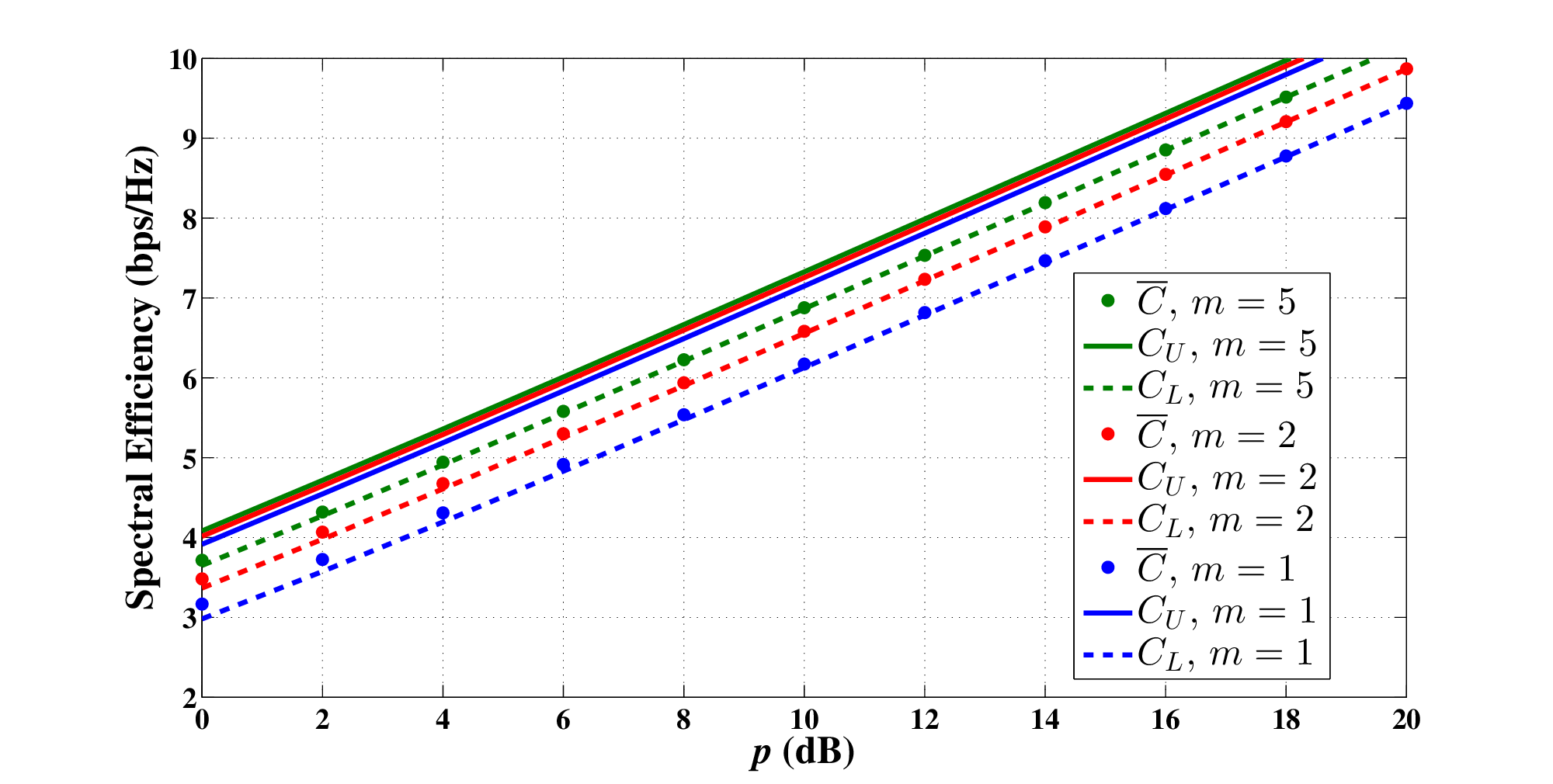}
\caption{Spectral efficiency vs. various values of the transmit SNR, where $N=1$, $\{M_{0},M_{1}\}=4$, and $\kappa=8$.}
\label{fig4}
\end{figure}

In Fig.~\ref{fig3}, the impact of imperfect CSI on the system performance is illustrated when the operators are equipped with large RIS arrays; a broader phase misalignment (i.e., lower $\kappa$) results on a system performance degradation. Moreover, the lower and upper bounds ($C_{L}$ and $C_{U}$, respectively) are tight in the large RIS regime and tend to be even more sharp for less CSI imperfection (higher $\kappa$). 
In Fig.~\ref{fig4}, the system performance is illustrated for small RIS arrays (i.e., $2\times 2$), a limited CSI imperfection and various Nakagami channel fading environments. Spectral efficiency is increased for a higher $m$ parameter (i.e., less rich scattering, closer to a near-LoS signal propagation) and higher transmit SNR values, as expected. It is worthy to note that the lower bound $C_{L}$ (although more involved) remains tighter in comparison to $C_{U}$ even for small RIS arrays. 

\begin{figure}[!t]
\centering
\includegraphics[trim=2.5cm 0.0cm 2.5cm 1.0cm, clip=true,totalheight=0.35\textheight]{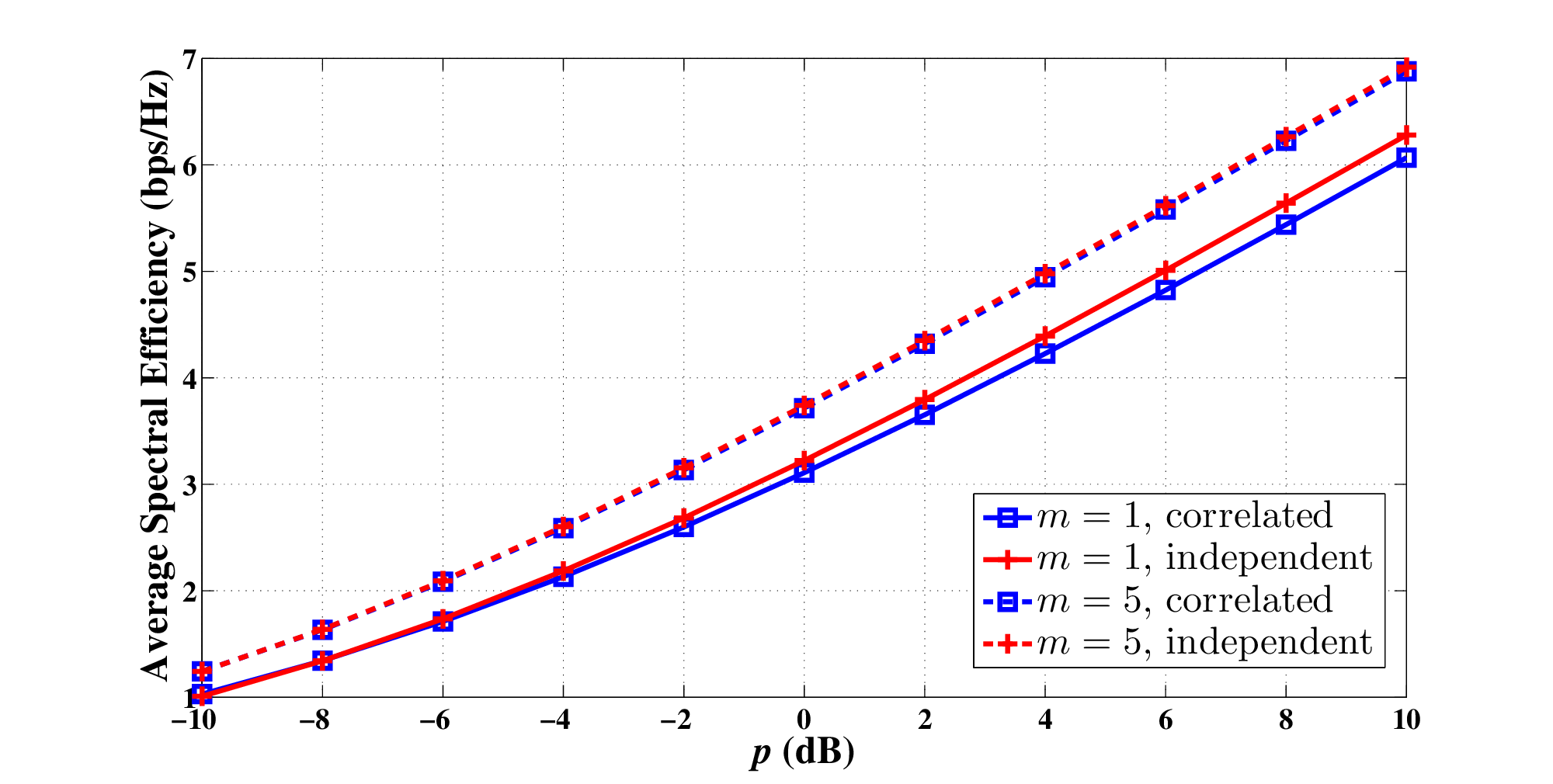}
\caption{Spectral efficiency vs. various values of the transmit SNR under spatially correlated or independent RISs, where $N=1$, $\{M_{0},M_{1}\}=4$, and $\kappa=8$.}
\label{fig5}
\end{figure}

In Fig.~\ref{fig5}, the impact of spatial correlation at RIS arrays is indicated on the average spectral efficiency $\overline{C}$. Obviously, the correlated (practical) scenario lower bounds the spectral efficiency of the (ideal) independent one. Notably, when signal propagation approaches LoS conditions, the system performance gap between the two cases becomes marginal. Doing so, spatial correlation at RISs seem not to play a critical role onto the system performance; especially when the channel fading conditions approach LoS-like and/or less scattering signal propagation.

\begin{figure}[!t]
\centering
\includegraphics[trim=2.0cm 0.0cm 2.5cm 1.0cm, clip=true,totalheight=0.35\textheight]{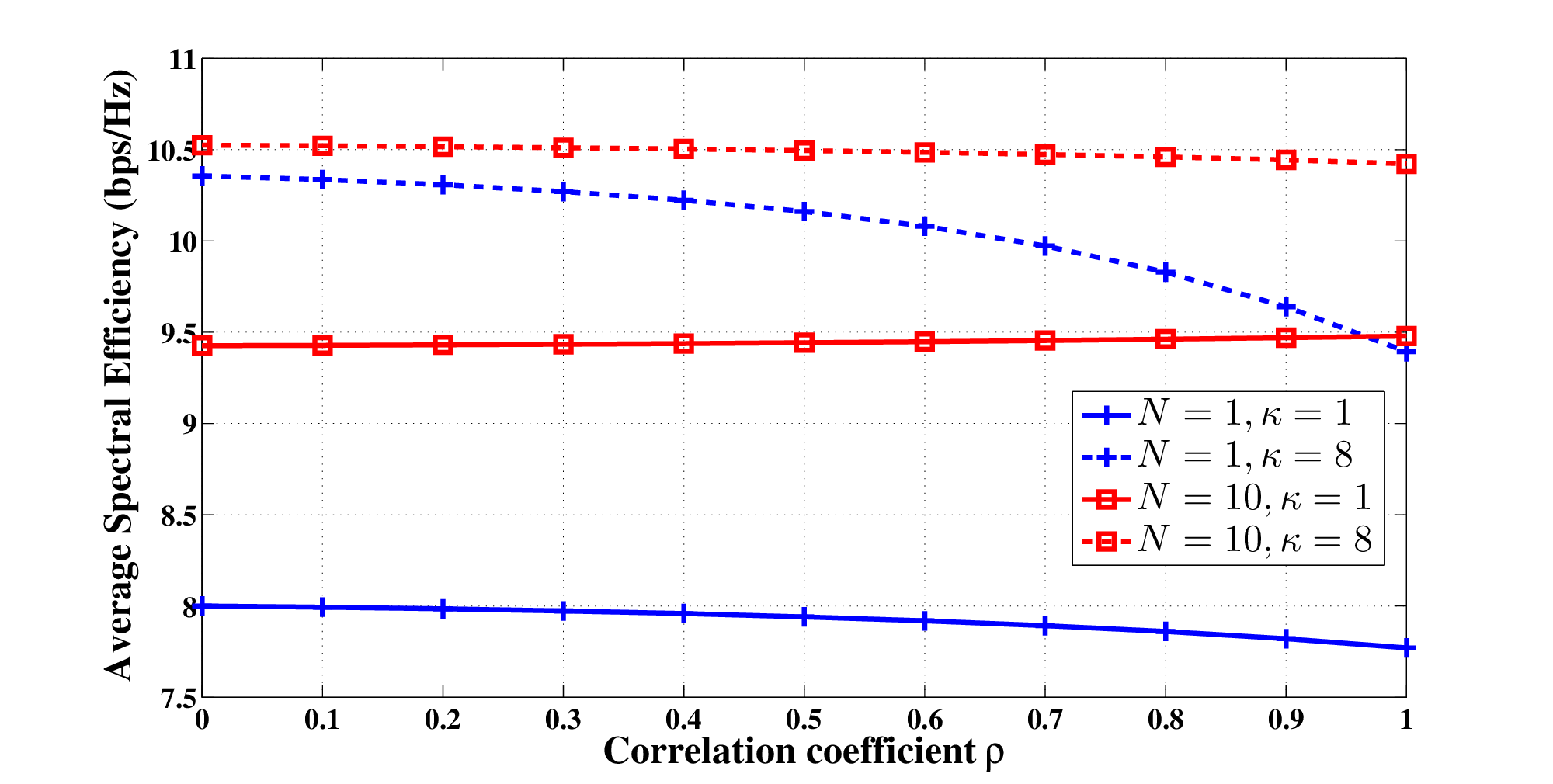}
\caption{Spectral efficiency vs. various values of the power correlation coefficient $\rho$ under a multi-operator environment and different phase mismatch concentrations, where $p=10$ dB, $\{M_{0},M_{1}\}=16$, and $m=1$.}
\label{fig6}
\end{figure}

{\color{black}Finally, Fig.~\ref{fig6} illustrates the average spectral efficiency of the considered system for several values of the power correlation coefficient at RIS (which in turn reflects on a different inter-element spacing). This system setup is evaluated over a lower (when $\kappa=8$) and higher ($\kappa=1$) phase mismatch. Also, the case when the reference system operates under the presence of a single IOI source is compared with the case when ten distinct IOI sources coexist ($N=10$ may seem a rather exaggerated use case in practice, yet it aims to better manifest the appearing trend). Specifically, a set of useful outcomes can be extracted as follows. Firstly, as previously verified in Fig.~\ref{fig5}, spatial correlation at RIS does not influence much the system performance; the worst-case scenario ($N=1,\kappa=8$) presents roughly a $1$ dB performance gap, from purely independent ($\rho=0$) to fully-correlated ($\rho=1$) channels. Insightfully, the presence of an increased number of distinct IOI sources (i.e., a higher $N$) reflects on a performance improvement. This occurs because more distinct IOI sources result to a richer scattered surrounding (i.e., enriched degrees-of-freedom for the received signal), which in turn enhances the system performance.} 

\section{Concluding Remarks}
The effect of IOI in a multi-operator/multi-RIS communication system was analytically studied. Spatial correlation at RIS arrays was considered, which is a practical condition in realistic applications. To cope with this, the versatile correlated Nakagami-$m$ channel fading model was adopted. Further, the various distinct IOI sources (originated from different operators) were modeled by random durations with respect to the time-frame of the reference system. This is a reasonable assumption due to the heterogeneous nature of distinct IOI sources within a multi-operator environment. The provided analysis produced tight and quite efficient lower and upper bounds of the system spectral efficiency as well as a corresponding asymptotic expression in the high RIS array regime. The derived results showcase that spatial correlation does not play a crucial role on the system performance under the presence of IOI, regardless of the RIS array size. This observation is even more emphatic whenever near-LoS signal propagation and/or less scattering occurs for the RIS-enabled links. 

\appendix
\label{appendix}
\numberwithin{equation}{section}
\setcounter{equation}{0}
Let ${\rm T}_{\rm d}=c_{i}\times {\rm T}_{{\rm d},i}$, where $c_{i}$ denotes the (integer) number of times ${\rm T}_{{\rm d},i}$ transmission intervals (regarding the $i^{\rm th}$ operator) are realized within ${\rm T}_{\rm d}$ or slightly exceeding it. First, we assume $N=1$ and we will extend the approach to $N>1$ operators afterwards. Then, normalizing the time interval for simplicity such that ${\rm T}_{\rm d}=1$ (and hence ${\rm T}_{{\rm d},1}=1/c_{1}$), the received signal of \eqref{received} becomes: ${\rm r}=\sqrt{\rm p}(\mathbf{h}^{T}_{0}\mathbf{\Phi}_{0}\mathbf{g}_{0}+\sum^{c_{1}-1}_{l=1}\sqrt{\frac{1}{c_{1}}}\mathbf{h}^{T}_{1}\mathbf{\Phi}_{1,l}\mathbf{g}_{1}+\sqrt{q_{1}}\mathbf{h}^{T}_{1}\mathbf{\Phi}_{1,c_{1}}\mathbf{g}_{1})s+w\approx \sqrt{\rm p}(\mathbf{h}^{T}_{0}\mathbf{\Phi}_{0}\mathbf{g}_{0}+\sum^{c_{1}-1}_{l=1}\sqrt{\frac{1}{c_{1}}}\mathbf{h}^{T}_{1}\mathbf{\Phi}_{1,l}\mathbf{g}_{1})s+w,$ $c_{1}\geq 2$, where the last approximation is due to the fact that the duration $q_{1}$ is nearly zero for practical applications (e.g., it is less than $10\mu s$ in 5G NR since each OFDM symbol may span from $8.3\mu s$ to $66.7\mu s$). Setting $\mathcal{Y}_{1}\triangleq e^{-{\rm j} \angle[\mathbf{h}^{T}_{0}\mathbf{\Phi}_{0}\mathbf{g}_{0}]}\sum^{c_{1}-1}_{l=1}\sqrt{1/c_{1}}\mathbf{h}^{T}_{1}\mathbf{\Phi}_{1,l}\mathbf{g}_{1}$, we easily deduce that $\mathbb{E}[\mathcal{Y}_{1}]=0$ and ${\rm Var}[\mathcal{Y}_{1}]=M_{1}(\frac{c_{1}-1}{c_{1}})$ with $M_{1}$ being the number of RIS elements of operator-1. Notice that the latter variance is bounded from $M_{1}/2$ (when $c_{1}=2$) to $M_{1}$ (when $c_{1}\rightarrow \infty$). Extending the above analysis to $N\geq 1$ IOI sources, we get ${\rm Var}[\sum^{N}_{i=1}\mathcal{Y}_{i}]=\sum^{N}_{i=1}M_{i}(\frac{c_{i}-1}{c_{i}})$. The said variance ${\rm Var}[\sum^{N}_{i=1}\mathcal{Y}_{i}]$ is nested in $\sigma^{2}$ defined below \eqref{CDFSNRcon}, whereas in this certain setup it becomes $\sigma^{2}\triangleq \sum^{N}_{i=1}\frac{M_{i}(c_{i}-1)}{2 c_{i}}$. All in all, we infer from Eqs. (14) and (19) that an increased $\sigma^{2}$ results in a higher lower and upper spectral efficiency bounds, correspondingly. Thereby, the average spectral efficiency is also increased for an increased $\sigma^{2}$, which in turn makes the worst-case scenario the one presented in the main manuscript. This outcome is reasonable since richer scattering channel conditions (i.e., an increased $c_{i}$) results to an enhanced system performance.


\bibliographystyle{IEEEtran}
\bibliography{IEEEabrv,References}

\vfill

\end{document}